\documentclass[prl,twocolumn,superscriptaddress]{revtex4-1}

\usepackage{amssymb}
\usepackage{graphicx}
\usepackage{dcolumn}
\usepackage{bm}
\usepackage{amsmath}
\usepackage{mathrsfs}

\usepackage{textcomp}

\usepackage[unicode=true,bookmarks=true,bookmarksnumbered=false,bookmarksopen=false,breaklinks=false,pdfborder={0 0 1},backref=false,colorlinks=true]{hyperref}

\hypersetup{linkcolor=magenta,urlcolor=blue,citecolor=blue,pdfstartview={FitH},hyperfootnotes=false,unicode=true}

\def\be{\begin{equation}}
\def\ee{\end{equation}}
\def\bea{\begin{eqnarray}}
\def\eea{\end{eqnarray}}
\def\nn{\nonumber}

\begin{document}
\title{Rotation symmetry enforced coupling of spin and angular-momentum for $p$-orbital bosons}
\author{Yongqiang Li}
\email{li\_yq@nudt.edu.cn}
\affiliation{Department of Physics, National University of Defense Technology, Changsha 410073, P. R. China}
\affiliation{Department of Physics, Graduate School of China Academy of Engineering Physics, Beijing 100193, P. R. China}
\author{Jianmin Yuan}
\affiliation{Department of Physics, National University of Defense Technology, Changsha 410073, P. R. China}
\affiliation{Department of Physics, Graduate School of China Academy of Engineering Physics, Beijing 100193, P. R. China}
\author{Andreas Hemmerich}
\affiliation{Institut f\"ur Laser-Physik, Universit\"at Hamburg, Luruper Chaussee 149, 22761 Hamburg, Germany}
\author{Xiaopeng Li}
\email{xiaopeng\_li@fudan.edu.cn}
\affiliation{State Key Laboratory of Surface Physics, Institute of Nanoelectronics and Quantum Computing, and Department of Physics, Fudan University, Shanghai 200433, China}
\affiliation{Collaborative Innovation Center of Advanced Microstructures, Nanjing 210093, China}

\begin{abstract}
{Intrinsic spin angular-momentum coupling of an electron has a relativistic quantum origin with the coupling arising from charged-orbits, which does not carry over to charge-neutral atoms.
Here we propose a mechanism of spontaneous generation of spin angular-momentum coupling with spinor atomic bosons loaded into $p$-orbital bands of a two-dimensional optical-lattice.
This spin angular-momentum coupling originates from many-body correlations and spontaneous symmetry breaking in a superfluid, with the key ingredients attributed to spin-channel quantum fluctuations and an approximate rotation symmetry. The resultant spin angular-momentum intertwined superfluid has Dirac excitations.
In presence
of a chemical potential difference for adjacent sites,
it provides a bosonic analogue of a symmetry-protected-topological insulator.
Through a dynamical mean-field calculation, this novel superfluid is found to be a generic low-temperature phase, and it gives way to Mott localization only at strong interactions and even-integer fillings. We show the temperature to reach this order is accessible with present experiments.
}

\end{abstract}

\date{\today}


\maketitle

{\it Introduction.---}
The interplay of spin, orbital, and charge degrees of freedom is one of the cornerstones of correlated quantum materials~\cite{2000_Tokura_Science}. Many intriguing quantum phases of electronic matter can be attributed to higher order electronic orbitals and spin-orbital interactions, for example, in exotic superconductivity in pnictides~\cite{2006_Kamihara_JACS} and strontium ruthenates~\cite{1998_Luke_Nature}, as well as in various topological insulators~\cite{2005_Kane_PRL,2007_Konig_Science} and Weyl semimetals~\cite{2011_Burkov_PRL}.

In ultracold atoms spectacular progress  has been made in the last decade  in controlling, and cooling various atomic species, paving a novel way for artificial quantum engineering of exotic atomic matter~\cite{2007_Lewenstein_AP,2008_Bloch_Dalibard_RMP,2010_Esslinger_CMP,2015_Lewenstein_RPP,2016_Li_Liu_RPP}. Fascinating  quantum many-body physics with ultra-low entropy~\cite{2016_Greiner_Nature} has been achieved in experiments of emulating  Bose- and Fermi-Hubbard models~\cite{1998_Zoller_Jaksch_PRL,2002_Hofstetter_Cirac_PRL} by loading atoms in the lowest band of optical lattices. To investigate the physical interplay of orbital degrees of freedom and charge (charge refers to atom number in charge neutral atomic systems), $p$-orbital systems have been explored extensively in both theory and experiments in recent years~\cite{lewenstein2011optical,2016_Li_Liu_RPP,Hemmerich11,Hemmerich2013,2013_Zhou_PRA,2016_Zhou_PRA}. Even this orbital-only  ``plain vanilla"  system without spin has already lead to a rich variety of correlated quantum effects. A fluctuating quantum orbital  liquid has been proposed~\cite{2010_Sun_PRL}, topological Chern insulators and superfluids  have been engineered by considering multi-orbital settings~\cite{li2013topological,2014_Liu_NC,2015_Lobos_PRX}, and a time-reversal symmetry breaking $p+ip$ Bose-Einstein condensate has been found in experiments~\cite{Hemmerich11,Hemmerich2013,kock2016orbital}.

The $p$-orbital atomic physics so-far established suggests that further introducing spin degrees of freedom  would potentially open up a new dimension for quantum engineering in optical lattices. Of particular interest is to explore the coupling between spin and $p$-orbitals~\cite{2016_Congjun_PRA}. It is well known that spin-orbital coupling is a necessary ingredient in the single-band $s$-orbital setting in order to enable topological phenomena and spin-Hall physics~\cite{1971_Yakonov_SJETPL}, which has been the reason for enormous efforts made in synthesizing artificial spin-orbit coupling using laser-assisted Raman coupling schemes ~\cite{2005_Osterloh_PRL,2005_Ruseckas_PRL,2009_Liu_PRL,2010_Dalibard_PRA,2011_Dalibard_Gerbier_RMP,2011_Lin_Nature,2012_Galitski_PRL,2012_Zhang_PRL,2012_Zwierlein_PRL,2013_Galitski_Nature,2015_Zhai_RPP,2015_Engels_PRL,2016_Chen_Science,2017_Sun_arXiv,2017_Liu_arXiv}.
The combination of spin and $p$-orbitals raises the intriguing question: {\it can spinor bosons in $p$-orbital bands give rise to spin-orbit coupled physics without engineering artificial spin-orbital coupling?}

The answer we provide here is positive.
In this work, we study two-component spinor bosons (spin referring to atomic hyperfine states of ultracold atoms) loaded into the $p$-orbital bands of a two-dimensional (2d) optical lattice, and establish a mechanism of spontaneous generation of spin angular-momentum coupling in a ground state superfluid. The resultant low-temperature phase is a spin angular-momentum intertwined (SAI) superfluid. Formulating a compact lattice Hamiltonian through symmetry analysis, the key to stabilize the  SAI order is found to be the interplay of an approximate rotational symmetry of the interaction and spin-channel quantum fluctuations. We argue that the SAI order could be further strengthened by including higher orbital bands. To incorporate finite interaction effects, a dynamical mean-field theory calculation is carried out. It is confirmed in the numerics that the SAI  superfluid is a generic low-temperature phase for the $p$-orbital spinor system, except that it undergoes a Mott localization at strong enough interactions and even-integer fillings. A complete phase diagram is mapped out to guide future experiments. The temperature to reach the SAI order is found to be well within the reach of state-of-the-art experiments.


\begin{figure*}[htp]
\includegraphics[trim = 0mm 0mm 0mm 0mm, clip=true, width=.9\linewidth]{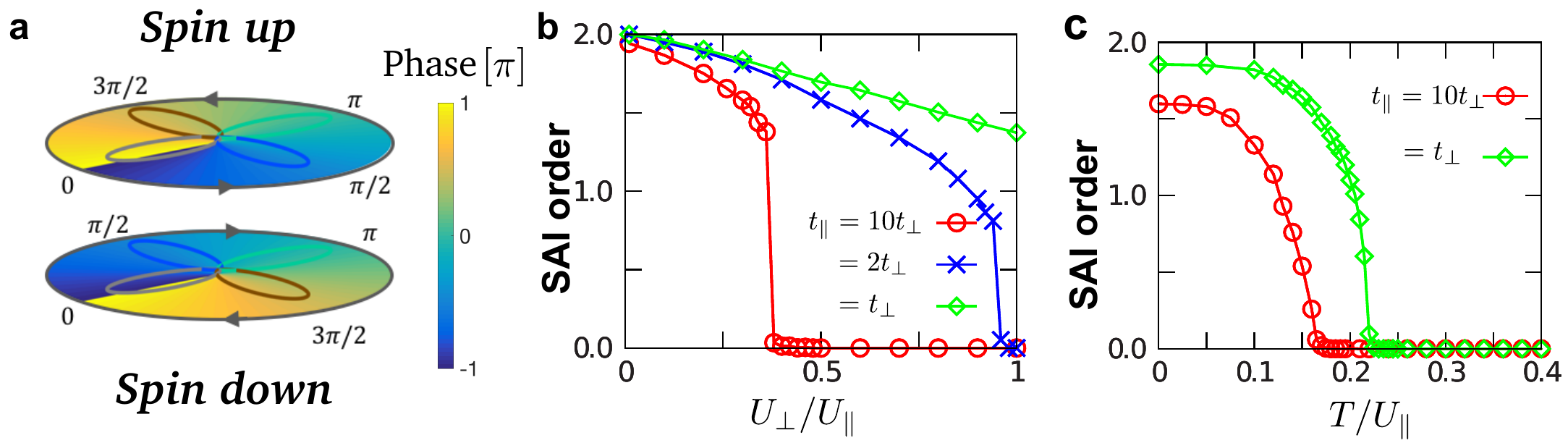}
\caption{(Color online) Rotation symmetry enforced spin angular-momentum intertwined (SAI) order: ({\bf a}) Pictorial illustration of SAI order.  In presence of the SAI order, the phase winding of spatial wave-function is entangled with the internal degrees of freedom of an atom in each optical lattice site.   (b) shows the stability of SAI order against interaction-driven quantum fluctuations. Here we fix the ratio between parallel tunneling of $p$-orbitals and the intra-spin interaction, $t_\parallel/U_{\parallel}=0.1$ (see Eq.~\eqref{eq:Ham}), and check the dependence of the SAI order on the inter-spin interaction, and the transverse tunneling   by varying $U_\perp/U_\parallel$, and $t_\parallel/t_\perp$. The phase-separation region with $U_\perp /U_\parallel >1$  is not shown in this plot. The SAI order parameter is introduced in Eq.~\eqref{eq:SAIorder}.
 ({\bf c}) Stability of SAI order against thermal fluctuations. In ({\bf c}), we  choose $U_\perp= 0.3 U_\parallel$, and $t_\parallel =  0.1U_\parallel$, and calculate the temperature dependence of the SAI order for $t_\perp/t_\parallel = 0.1$, and $1$.
The atomic filling in this figure is set to be $\langle n\rangle = 2$.
}
\label{fig_order}
\end{figure*}

\medskip
{\it Spinor bosons in a 2d $p$-orbital lattice.---}
The model Hamiltonian describing two-component spinor bosons in continuous space reads
$
H = \int d^3{\bf x}
 \left\{  \Phi^\dag \left[ -\frac{\hbar^2}{2M} \vec{\nabla}^2 \right] \Phi  \right.
			+ \frac{1}{2}
\left.
\left[ c_0  :\Phi^\dag \Phi \Phi^\dag \Phi:  + c_2 : \Phi^\dag \sigma_z \Phi \Phi ^\dag \sigma_z \Phi: \right] \right\}
$,
with $\Phi$ a two-component spinor field, $M$ the atomic mass, and $:\ldots:$ referring to normal ordering~\cite{1998_Ho_PRL,1998_Ohmi_JPSJ}.  The interaction strengths in charge and spin channels are characterized by $c_0$ and $c_2$. The ratio $c_2/c_0$ is tunable by using different atomic species, or by exploiting a Feshbach resonance~\cite{Widera2004Entanglement, B.Gadway}.
In this work, we focus on the spin-miscible case with $c_2/c_0>0$~\cite{miscible1, miscible2, miscible3}.
Consider the spinor bosons loaded into the $p$-orbital bands of a 2d square optical lattice, as in the experiments~\cite{Hemmerich11, Hemmerich2013,supplement} (the spinless case has been carried out in these experiments).
The low-energy physics of this system is described by a spinful $p$-orbital lattice model~\cite{supplement},
\bea
\label{eq:Ham}
H &=& \sum_{\bf r, \mu \nu }\left[ t_\parallel \delta_{\mu \nu} -t_\perp (1-\delta_{\mu \nu} )  \right]
			\left(\Phi_{\mu, {\bf r}} ^\dag  \Phi_{\mu,{\bf r}   +  {\bf e}_\nu}  + h.c. \right)  \nn \\
&+& \frac{U_0}{2} \sum_{\bf r}   \left[ \frac{2}{3} : n^2 : -\frac{1}{3} : L_z^2 :  + \frac{1}{3} :{\bf S} ^2: \right]    \\
&+& \frac{U_2}{2}  \sum_{\bf r}  \left[\frac{1}{3} :n^2: -\frac{1}{3} : LS_{z} ^2: + (:S_z ^2: - \frac{1}{3} :{\bf S} ^2:)  \right]. \nn
\eea
Here the lattice annihilation operators $\Phi_{x, {\bf r}}    $ and $\Phi_{y, {\bf r}}$ correspond to $p_x$- and $p_y$-orbitals at the lattice site ${\bf r}$ having localized Wannier wave-functions $w_x$ and $w_y$. The unit vectors ${\bf e}_\nu$ denote the primitive  lattice vectors, the interaction strengths are  $U_{m=0,2} =  c_m \int d^3 {\bf x} w_\mu^4 $, and
the rotational invariants include
number density $n = \sum_\mu \Phi^\dag_\mu \Phi_\mu$,
angular momentum $L_z=[i\Phi_y ^\dag \Phi_x + h.c. ]$,
spin moment ${\bf S} = \sum_\mu \Phi^\dag_\mu \vec{ \sigma} \Phi_\mu $, and a spin angular-momentum coupled operator,
\be
\textstyle {LS}_{z} = [i\Phi_y ^\dag \sigma_{z} \Phi_x + h.c. ].
\label{eq:SAIorder}
\ee
For convenience, we also introduce $U_\parallel = U_0 + U_2$, and $U_\perp = U_0 - U_2$, which correspond to intra- and inter-spin interactions, respectively. The compact interaction form obtained relies on the local rotation symmetry which approximately holds for deep optical lattices. Including rotation asymmetric interactions will induce additional complication for our theory analysis but is {\it not} expected to affect our results qualitatively as long as the symmetry-broken terms are not too strong.

{\it Spontaneous spin angular-momentum coupling.---}
In absence of inter-spin interaction between the two components, i.e., $U_0 = U_2$ or equivalently $U_\perp = 0$.
The spinor Hamiltonian reduces to two spin-decoupled copies of $p$-orbital Bose-Hubbard models whose phase diagrams have been studied extensively in both theory and experiments~\cite{Liu2006, Wu2006, Cai2013, Hemmerich2013}.  The ground-state at weak interaction is established to be a chiral $p_x \pm ip_y$ Bose-Einstein condensate. Introducing angular-momentum carrying operators
$\tilde{\Phi}_{\pm} ({\bf q})
	= \sum_{\bf r} e^{i {\bf q} \cdot {\bf r}}
	\left [ (-)^{r_x}\Phi_{x,{\bf r}}  \pm i (-) ^{r_y} \Phi_{y, {\bf r}} \right] $, there are four orthogonal degenerate ground-state condensates
$
1/\sqrt{N_\uparrow! N_\downarrow!} {[\tilde{ \Phi} _{\pm \uparrow} ^\dag ( 0) ]^{N_\uparrow} } {[\tilde{ \Phi} _{\pm \downarrow} ^\dag ( 0) ]^{N_\downarrow} } | 0\rangle,
$
with $|0\rangle $ referring to the vacuum. These  states can be specified by their total $L_z$ and  ${LS_z}$ quantum numbers $(m_l, m_{ls} )$.
Here we focus on spin-balanced case $N_\uparrow = N_\downarrow = N/2$, which is a stable configuration under spin-miscible condition.
The  four condensates  are denoted as
$|\chi^c_\pm \rangle$,  $|\chi^s_\pm \rangle$ according to
\bea
&&\left[\sum_{\bf r} (-1)^{r_x + r_y} L_z \right]| \chi^c _\pm \rangle = \pm N | \chi^c _\pm \rangle , \nn \\
&&\left[\sum_{\bf r} (-1)^{r_x + r_y} {LS_z}\right] |\chi^s _ \pm \rangle = \pm N | \chi^s _\pm \rangle . \nn
\eea
The alternating signs account for the fact that the $p$-orbital band minima are located at the Brillouin zone boundaries~\cite{Liu2006}. The states $|\chi^c _\pm\rangle$ are related to each other by  time-reversal symmetry (${\cal T}$), and must have the same energy at all perturbative orders even when we consider a finite inter-spin interaction $U_\perp\neq0$. The  states $|\chi^s_\pm \rangle $ are related by a combined symmetry of time-reversal with a spin flip (${\cal T}\rtimes \sigma_x$), and their degeneracy also survives at all orders. But  the mutual degeneracy between $|\chi^c \rangle$ and $|\chi ^s \rangle$ is lifted considering second order perturbations in $U_\perp$ owing to the difference in the following  commutation relations
\be
[L_z, {\bf S} ^2] = 0; \,\,\,\,\, \,\,\,\,\,\,\,
[{LS_z}, {\bf S}^2 ] \neq 0.
\ee
Interactions between two spin components for the case of $|\chi_c\rangle $ cannot cause spin exchange due to angular momentum $L_z$ conservation, whereas interactions for $|\chi_s\rangle $ do cause spin exchange because the spin angular-momentum  coupling is not conserved. Taking $\chi_{c+}$ and $\chi_{s+}$ for an example, the former only couples to
$\Phi_{+\uparrow} ({\bf q}) \Phi_{+ \downarrow} ^\dag (-{\bf q}) \Phi_{+ \uparrow} (0) \Phi_{+\downarrow} (0) | \chi_{c+} \rangle$, whereas the latter couples to two types of states
$ \Phi_{\pm \uparrow} ({\bf q}) \Phi_{\mp \downarrow} ^\dag (-{\bf q}) \Phi_{+ \uparrow} (0) \Phi_{-\downarrow} (0) | \chi_{s+} \rangle$. From the spin-exchange quantum fluctuations, the
$|\chi_{s} \rangle$ states receives an additional energy correction $\Delta E$ given as,
\be
\Delta E/N_s = -\frac{1}{9} \rho^2 U_\perp ^2 \int \frac{d ^2 {\bf q}}{(2\pi)^2 } \frac{ \epsilon_{\bf q}  }{(\epsilon_{\bf q}^2 - z_{\bf q}^2 )},
\ee
with $N_s$ the number of lattice sites, $\rho$ the average total atomic number density.
Here we introduced
$\epsilon_{\bf q} = -2(t_\parallel+t_\perp) (\cos q_x + \cos q_y -2)   $,
$z_{\bf q} = (t_\perp - t_\parallel) (\cos q_x - \cos q_y)$.

With perturbative spin interactions, i.e., $U_\perp =  U_0 -U_2$ being small, we have thus established that the $|\chi_s\rangle$ condensate is the true ground state. In this state, we have spontaneous spin angular-momentum coupling, i.e., $\langle {LS_z} \rangle = \pm \rho$. This spontaneous order is staggered in real space. We ``dub" this novel state a spin angular-momentum intertwined superfluid. A pictorial illustration of the SAI order is shown in Fig.~\ref{fig_order}({\bf a}).  On each lattice site, the atomic orbital angular momentum is locked with regard to the spin moment. More precisely, the atoms with spin up and spin down acquire  local orbital $p_x + ip_y$ and $p_x - ip_y$ wave functions, respectively, or vice versa. Based on the Chern-insulator like Bogoliubov energy spectra obtained for the single spin-component case~\cite{2015_Engelhardt_PRA,2015_Ueda_NJP,Zhifang2016,2016_Morais_PRL,2018_Lu_arXiv}, the SAI superfluid is expected to exhibit Dirac excitations. Furthermore, if a potential imbalance for neighboring sites is implemented with superlattice techniques, this superfluid would  provide a bosonic analogue of time-reversal symmetry protected electronic topological insulators with edge modes that carry chiral spin currents. The topological excitation spectra are provided in Supplementary Material.

\begin{figure}
\includegraphics[width=.45\textwidth]{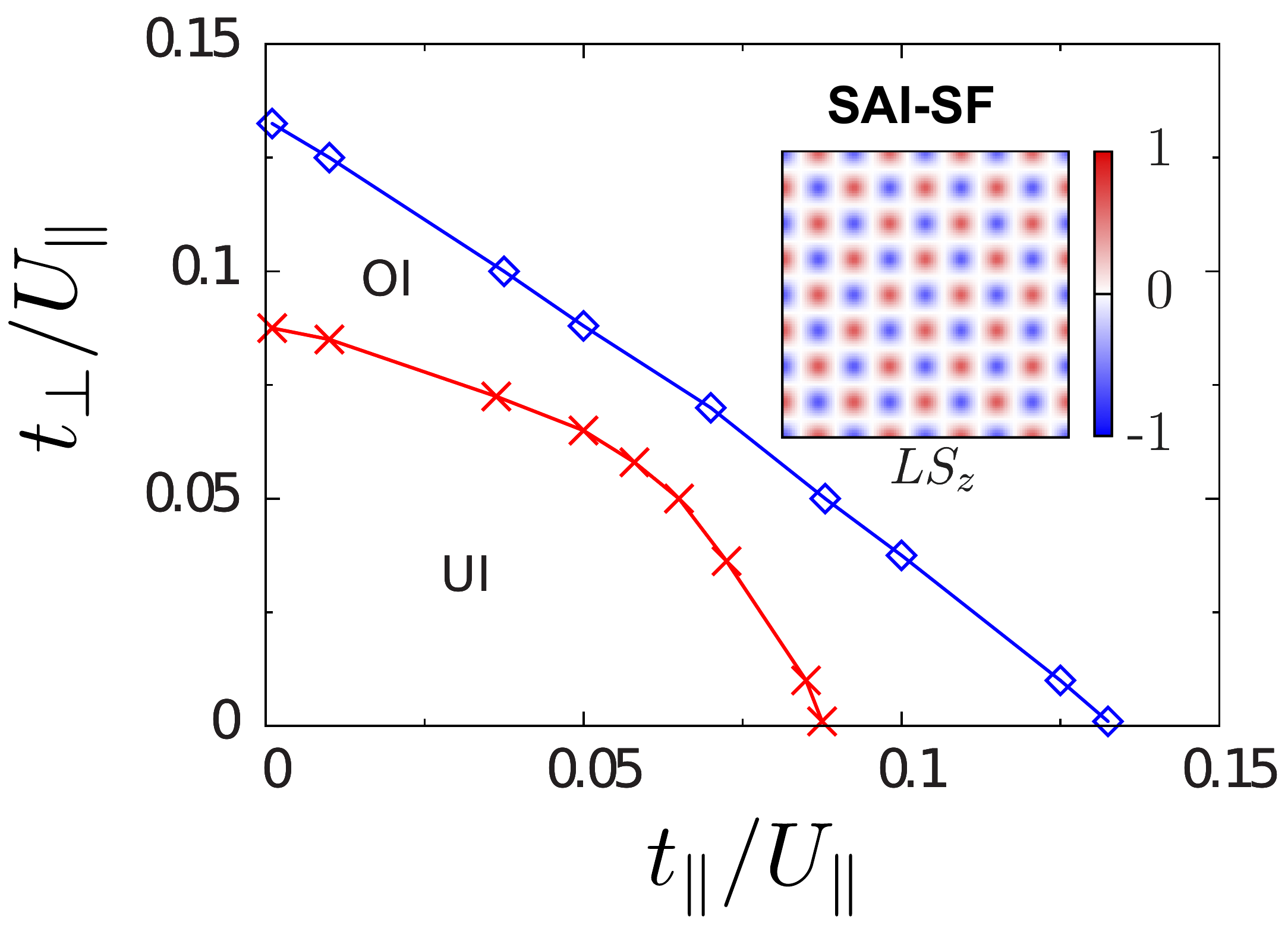}
\caption{(Color online) Phase diagram of the spinful $p$-orbital system with an even integer filling.  The phase diagram is obtained via bosonic dynamical mean-field theory. The atomic filling, i.e. number of particles per lattice site, is fixed at  $\langle n\rangle=2$. Varying $t_\parallel/U$ and $t_\perp/U$, we find three phases in the system, including  the SAI superfluid together with two Mott states---unordered insulator (UI) and an ordered insulator (OI).  The spin order $\langle {\bf S}^2 \rangle$ is vanishing (finite) for UI (OI), which differentiates the two Mott states. The inset shows the the staggered spin angular-momentum intertwined order in real space in the {\it SAI}-SF. In this plot, we use $U_\parallel = U_\perp$.}
\label{fig_t}
\end{figure}

{\it Stability of SAI order against strong interactions and thermal fluctuations.---}
To show the robustness of SAI order against quantum and thermal fluctuations caused by strong interactions and finite temperature, we go beyond the perturbative treatment and  carry out a bosonic dynamical mean-field (BDMFT) calculation~\cite{Hubener, Tong, Werner, Li2011,Li2012,Li2013,Liang15,Li2016, Li2017}. In order to account for the inhomogeneous order on the lattices, a real-space four-component BDMFT is  implemented~\cite{supplement}.
The reliability of this approach has been confirmed by means of a comparison with an unbiased quantum Monte-Carlo simulation~\cite{QMC_boson}.

As shown in Fig.~\ref{fig_order}({\bf b}), the SAI order is observed to be stable against  strong interactions. As we increase the inter-spin interactions $U_\perp$, while the SAI order becomes weaker, it remains stable even for moderate interaction strength. When $U_\perp$ is too strong, we find a first-order transition where the SAI order  vanishes in an abrupt fashion. We also looked at the dependence on the ratio $t_\perp/t_\parallel$, which is tunable in experiments by controlling the lattice geometry~\cite{Hemmerich11}. Our results show that the SAI order becomes stronger as we increase the $p$-orbital transverse tunneling $t_\perp$, and gets maximally stable for $ t_\perp = t_\parallel$. A related scenario can be realized in a bipartite lattice geometry as implemented in Ref.~\cite{Hemmerich11,2016_Li_Liu_RPP}.

In order to determine the temperature window to reach SAI order, we investigate its finite temperature phase transition. The results are shown in Fig.~\ref{fig_order}({\bf c}). The finite temperature phase transition is found to be continuous. For both parameter choices of $t_\perp/t_\parallel$ $= 0.1$ (small) and $1$ (big), we find the critical temperature of the SAI order is to the same order of band width. We notice that the critical temperature of SAI order is just slightly below the superfluid transition in our system, which is another evidence of the  SAI order robustness.  The SAI superfluid is thus accessible to  present cold-atom experiments with standard evaporative cooling techniques. 

\begin{table}
    \caption{Characterization of different quantum phases for bosonic mixtures in the $p$-orbital bands in an optical lattice. The definition of these orders ($\Phi_{\mu}$, $S_z$, ${\bf S}^2$, $L_z$ ${{LS_z}}$) can be found in the main text. }
    \begin{tabular}{p{0.1\textwidth}|p{0.06\textwidth}p{0.06\textwidth}p{0.06\textwidth}p{0.06\textwidth}p{0.06\textwidth} p{0.056\textwidth}p{0.056\textwidth} } \hline \nonumber
    Phases         & $\langle \Phi_{\mu} \rangle $     		&$\langle S_z \rangle $		& $\langle {\bf S}^2 \rangle $      	&$\langle L_z\rangle $		 & $\langle {LS_{z}}\rangle $		\\ 
\hline
    {\it SAI$_z$}-SF   & $\neq 0$           	&$=0$	 	&  $\neq 0$   	&$=0$		& $\neq 0$  	 	\\ 
    OI             & $=0$               			&$=0$  		&  $\neq 0$    	&$=0$		& $= 0$   			\\
    UI             & $=0$                			&$=0$		&  $=0$		&$=0$		& $=0$       		\\
   \hline
    \end{tabular}
\label{tableI}
\end{table}

\begin{figure}
\includegraphics[width=\linewidth]{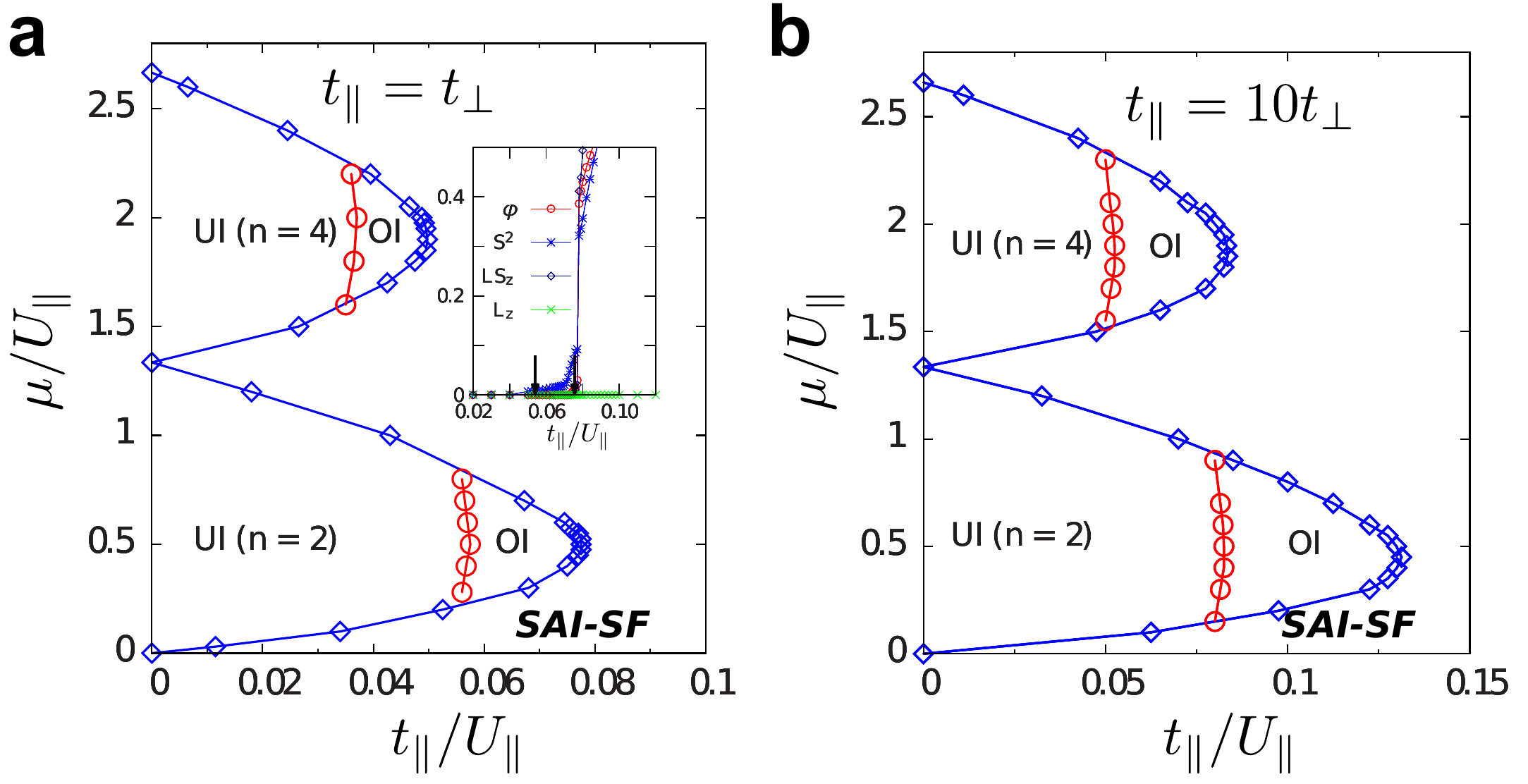}
\caption{(Color online) Phase diagrams of  spinful $p$-orbital bosons at generic fillings. ({\bf a}) and ({\bf b}) correspond to different tunneling ratios, $t_\parallel/t_\perp = 1$ and $10$, respectively. BDMFT predicts that the system supports various quantum phases, including unordered Mott insulator (UI), ordered Mott insulator (OI), superfluid with spin angular-momentum intertwined order ({\it SAI}-SF). The inset in ({\bf a}) shows the evolution of the order parameters  along $\mu/U_\parallel=0.6$.  We use interaction strengths,  $U_\parallel = U_\perp$.}
\label{fig_mu}
\end{figure}

{\it Phase Diagram.---} We further map out the phase diagram for the $p$-orbital spinor bosons (see  Hamiltonian in Eq.~\eqref{eq:Ham}) using BDMFT. The order parameters we check include superfluid order $ \Phi_{x,y,\sigma}$, magnetic orders $S_z$ and ${\bf S}^2$, orbital order $L_z$, and the SAI order ${{LS_z}}$.

Since an analogue of tuning the ratio $t_\parallel/t_\perp$ becomes possible  in experiments~\cite{Hemmerich11} using a bipartite lattice geometry, we provide a phase diagram parameterized by $t_\parallel/U_\parallel$ and $t_\perp/U_\parallel$ in Fig.~\ref{fig_t} where the atomic filling is fixed to be $\langle n\rangle = 2$. The SAI-superfluid is found to be a stable phase that occupies a large region of the phase diagram. Only when the interaction is strong enough to support a Mott transition, the SAI-superfluid would disappear. This suggests the SAI-superfluid is a generic ground state for weakly interacting spinor-bosons loaded in $p$-orbital bands.


We further investigate the dependence of the SAI order on atomic fillings, with numerical results shown in Fig.~\ref{fig_mu}. It is confirmed that the SAI-superfluid appears at generic atomic fillings. Only at even-integer fillings, the SAI superfluid develops a transition to Mott localization with strong interactions. The spin order phase transition in the Mott regime has similarity to spin-$1$ bosons in the lowest band~\cite{2003_Demler_PRA}. Other than that, the SAI-superfluid is found to be a  generic ground-state for the system.

Our numerical results confirm that the SAI superfluid is a generic low-temperature phase for interacting spinor-bosons loaded in $p$-orbital bands. It is expected to be as robust as the chiral $p$-wave condensate already observed experimentally~\cite{Hemmerich11,2014_Li_NC,2016_Li_Liu_RPP}. Taking our theoretical analysis, we attribute  the robustness of this novel phase to the protection from the local rotation symmetry.

We note here that the system has two different Mott states in the strong interaction region (see Figs.~\ref{fig_t}, and~\ref{fig_mu}). One is a featureless Mott insulator, and the other has a spin order $\langle {\bf S}^2\rangle \neq 0$. {The local state for the former is a spin-singlet formed by two orbitals, and it has overlap with spin-triplets for the latter.} It is yet worth remarking here that the Mott states in the $p$-orbital system are difficult to reach because the required strong interactions would dramatically enhance the decay process from the $p$-orbital to the ground band, which makes a lifetime too short  for the system to equilibrate to Mott localization.

\medskip
{\it Discussion.---}
There are two remarks we would like to make here. First, the SAI order is beyond the Gross-Pitaevskii (GP) mean-field theory. GP theory would not distinguish $|\chi_c\rangle$ (with $\langle LS_z \rangle =0$)  and $|\chi_s\rangle$ (with $\langle LS_z \rangle \neq 0$) states as they would be precisely degenerate. The SAI order appears solely from quantum interaction effects, which is the reason why the novel SAI order was missing in a previous GP-mean-field study~\cite{2016_Congjun_PRA}. Second, the SAI order would be further strengthened by considering higher band effects. If we incorporate cross-band fluctuations, the  argument based on angular momentum conservation to support the SAI superfluid still leads to ``softer" fluctuation phase space for  SAI ordered $|\chi_s \rangle$ states compared to $|\chi_c \rangle$. The SAI order is thus expected to be even more robust considering multi-band effects.

We emphasize here that the SAI superfluid is a generic stable phase for two-component spinor bosons loaded into $p$-orbital bands. This novel phase does not require any fine tuning. This has been established with our theoretical analysis and further supported by exclusive numerical results.  Furthermore its transition temperature is accessible with present experimental techniques~\cite{2016_Li_Liu_RPP,Hemmerich11,Hemmerich2013,2013_Zhou_PRA,2016_Zhou_PRA}.
Since spontaneous Ising symmetry breaking has been observed in atomic systems, e.g., in Refs.~\onlinecite{Struck2011Quantum,baumann2010dicke}, we expect the spontaneous spin angular-momentum order and its topological consequences to be plausible to experimental realization.

\medskip
\paragraph*{Acknowledgement.---}
We acknowledge helpful discussion with Shuai Chen, Sai-Jun Wu, Zhi-Fang Xu, Guang-Quan Luo, Jin-Sen Han. This work is supported by National Program on Key Basic Research Project of China under Grant No. 2017YFA0304204 (XL), National Natural Science Foundation of China under Grants No. 11304386 (YL), 11774428 (YL), 117740067(XL),  and the Thousand-Youth-Talent Program of China (XL). The work was carried out at National Supercomputer Center in Tianjin, and the calculations were performed on TianHe-1A. AH acknowledges support by the German Research Foundation (DFG) under grant SFB925-C2.

\clearpage
\begin{widetext}
\begin{center}
{\Huge \bf Supplementary Material}
\end{center}
\renewcommand{\theequation}{S\arabic{equation}}
\renewcommand{\thesection}{S-\arabic{section}}
\renewcommand{\thefigure}{S\arabic{figure}}
\renewcommand{\bibnumfmt}[1]{[S#1]}
\renewcommand{\citenumfont}[1]{S#1}
\setcounter{equation}{0}
\setcounter{figure}{0}

\section{Experimental protocol}
The present work considers the second band of a monopartite square lattice composed of local $p_x$- and $p_y$-orbitals. As will be discussed in detail in forthcoming work, the physics presented here also occurs in a bipartite square lattice, where shallow and deep wells are arranged as the black and white fields of a chequerboard. The deep wells host $p_x$- and $p_y$-orbitals, while the shallow wells provide $s$-orbitals. It has been shown in Ref.~\cite{Hemmerich11} that this lattice potential can be realized such that the depth of both type of sites can be rapidly switched in order to enable Landau-Zener dynamics with the result that the second band of the lattice can be efficiently populated. In a subsequent rethermalization process, chiral condensates have been produced showing lifetimes of nearly up to a second. This technique can be readily extended to the case of spinor condensates as well. Starting with an atomic sample prepared in a single Zeeman substate, the preparation of a two-component mixture can be conducted by adiabatic passage or application of a $\pi/2$ radio-frequency pulse, exploiting the quadratic Zeeman effect~\cite{2004_Krutitsky_PRA}. This can be achieved before or after the atoms have been excited to the second band. The SAI order could be identified by interference similarly as in Ref.~\cite{2015_Hemmerich_PRL} or by related protocols employing the spin degrees of freedom.

\section{Details of derivation for the spinful $p$-orbital Bose-Hubbard Model}
In this supplementary section, we provide details of the derivation for the spinful $p$-orbital Bose-Hubbard model in Eq.~(1). The starting point is the Hamiltonian in the continuous space,
\be
H = \int d^3{\bf x}
 \left\{  \Phi^\dag \left[ -\frac{\hbar^2}{2M} \vec{\nabla}^2 \right] \Phi  \right.
			+ \frac{1}{2}
\left.
\left[ c_0  :\Phi^\dag \Phi \Phi^\dag \Phi:  + c_2 : \Phi^\dag \sigma_z \Phi \Phi ^\dag \sigma_z \Phi: \right] \right\}.
\label{eq:SHam}
\ee
For spinor bosons loaded into the $p$-orbital bands of a deep optical lattice considered in this work, the cross-band fluctuations are greatly suppressed due to energy suppression. The low energy fluctuations are restricted to $p$-bands only. To capture such low-energy fluctuations, the field operator $\Phi({\bf x})$ can be rewritten in terms of lattice annihilation/creation operators as
\be
\Phi ({\bf x}) = \sum_{\bf r} w_x ({\bf x} - {\bf r}) \Phi_{x, {\bf r}}  + w_y ({\bf x } - {\bf r}) \Phi_{y, {\bf r}},
\ee
with $\Phi_{x, {\bf r}}    $ and $\Phi_{y, {\bf r}} $ the lattice operators corresponding to localized Wannier $p_x$- and $p_y$-orbitals residing on site ${\bf r}$, and  $w_{x,y}$ the Wannier functions for the $p_x$ and $p_y$ orbitals.
For a deep optical lattice, each lattice site has an approximate rotation symmetry, which implies
\be
\int d^3 {\bf x} w_x ^4 = \int d^3 {\bf x} [ \cos \theta w_x + \sin \theta w_y]^4
\ee
for arbitrary rotation angle $\theta$. Then immediately we have  $\int d^3 {\bf x} |w_x |^4 = 3 \int d^3 {\bf x} |w_x|^2 |w_y| ^2$.
Withe the local rotation symmetry, the interaction must only contain rotation invariants such as
$L_z^2$, $n^2$, $S_{x,y,z}^2$, and $LS_z ^2$. It is straightforward to show that
\bea
 n^2  &=& : \left[ (\Phi_x^\dag \Phi_x)^2 + (\Phi_y^\dag \Phi_y)^2\right]  + 2 (\Phi_x ^\dag \Phi_x )(\Phi_y ^\dag \Phi_y):,   \nn \\
L_z ^2 &=&
: 2 (\Phi_x ^\dag \Phi_y ) (\Phi_y^\dag \Phi_x )
- \left[ (\Phi_x^\dag \Phi_y)^2 + (\Phi_y ^\dag \Phi_x) ^2  \right]:, \nn \\
S_z ^2 & = &  : \left[ (\Phi_x^\dag \sigma_z \Phi_x)^2 + (\Phi_y^\dag \sigma_z \Phi_y)^2\right]  +2 (\Phi_x ^\dag  \sigma_z \Phi_x )(\Phi_y ^\dag \sigma_z \Phi_y):,  \nn \\
{\bf S} \cdot {\bf S} &=&
 :\left[ (\Phi_x^\dag \Phi_x)^2 + (\Phi_y^\dag \Phi_y)^2\right]  -2 (\Phi_x ^\dag \Phi_x )(\Phi_y ^\dag \Phi_y)
+ 4 (\Phi_x ^\dag \Phi_y ) (\Phi_y^\dag \Phi_x ):, \nn \\
LS_z ^2 &=& : 2 (\Phi_x ^\dag \sigma_z \Phi_y ) (\Phi_y^\dag \sigma_z \Phi_x ) -\left[ (\Phi_x^\dag \sigma_z \Phi_y)^2 + (\Phi_y ^\dag \sigma_z \Phi_x) ^2 \right]:.
\label{eq:SRinvariants}
\eea
One useful identity worth keeping in mind is
$  :(\Phi_x ^\dag \Phi_x )(\Phi_y ^\dag \Phi_y)
- (\Phi_x ^\dag \Phi_y ) (\Phi_y^\dag \Phi_x ): = :-(\Phi_x ^\dag \sigma_z \Phi_x )(\Phi_y ^\dag \sigma_z \Phi_y)
+ (\Phi_x ^\dag \sigma_z \Phi_y ) (\Phi_y^\dag \sigma_z \Phi_x ) :$ .

The interaction in the charge channel, i.e., the $c_0$ term in Eq.~\eqref{eq:SHam}, is rewritten as
$
\sum_{\bf r} \int d^3 {\bf x} c_0 : \left[ (w_x \Phi_{x, {\bf r}} ^\dag + w_y \Phi_{y, {\bf r}} ^\dag) (w_x \Phi_{x, {\bf r}}  + w_y \Phi_{y, {\bf r}} )  \right]^2:,
$
where off-site interactions are neglected because they are exponentially small in a deep lattice. Taking the rotation symmetry, this interaction term reduces to
$$
\frac{U_0}{2} \sum_{\bf r} \left\{ \left[ (\Phi_x^\dag \Phi_x)^2 + (\Phi_y^\dag \Phi_y)^2\right]  + \frac{2}{3} (\Phi_x ^\dag \Phi_x )(\Phi_y ^\dag \Phi_y)
+ \frac{2}{3} (\Phi_x ^\dag \Phi_y ) (\Phi_y^\dag \Phi_x )
+ \frac{1}{3} \left[ (\Phi_x^\dag \Phi_y)^2 + (\Phi_y ^\dag \Phi_x) ^2  \right]
 \right\} .
$$
Comparing this form with the rotation invariants in Eq.~\eqref{eq:SRinvariants}, the compact form shown in the main text is obtained.

The interaction in the spin channel, i.e., the $c_2$ term in Eq.~\eqref{eq:SHam}, is rewritten as
$
\sum_{\bf r} \int d^3 {\bf x} c_2 : \left[ (w_x \Phi_{x, {\bf r}} ^\dag + w_y \Phi_{y, {\bf r}} ^\dag) \sigma_z (w_x \Phi_{x, {\bf r}}  + w_y \Phi_{y, {\bf r}} )  \right]^2:.
$
Considering rotation symmetry, the $c_2$ term takes a form of
$$
\frac{U_2}{2}  \sum_{\bf r} \left\{ \left[ (\Phi_x^\dag \sigma_z\Phi_x)^2 + (\Phi_y^\dag \sigma_z \Phi_y)^2\right]
+ \frac{2}{3} (\Phi_x ^\dag \sigma_z \Phi_x )(\Phi_y ^\dag \sigma_z \Phi_y)
+ \frac{2}{3} (\Phi_x ^\dag \sigma_z  \Phi_y ) (\Phi_y^\dag \sigma_z\Phi_x )
+ \frac{1}{3} \left[ (\Phi_x^\dag  \sigma_z \Phi_y)^2 + (\Phi_y ^\dag \sigma_z \Phi_x) ^2  \right]\right\} .
$$
In terms of rotation invariants in  Eq.~\eqref{eq:SRinvariants}, the compact interaction form shown in the main text is reached.


\section{topological Bogoliubov excitations}

\begin{figure}[htp]
\includegraphics[trim = 0mm 0mm 0mm 0mm, clip=true, width=.7\linewidth]{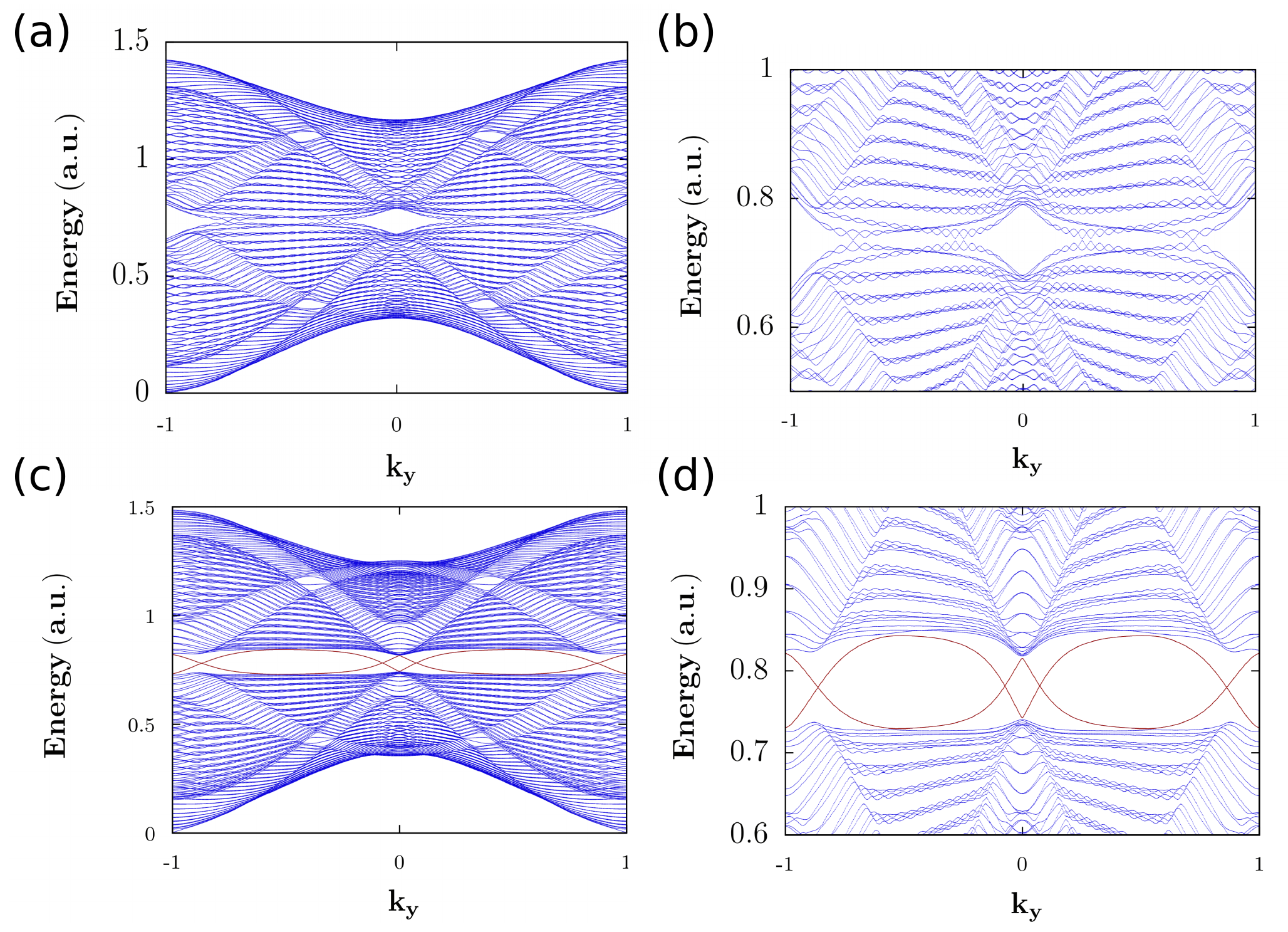}
\caption{(Color online) Topological excitations as a consequence of the SAI order. (a) and (c) show the excitations with and without potential difference between adjacent lattice sites. (b) and (d) show the enlarged view of (a) and (c), respectively, to highlight the topological excitations. Here we choose open boundary in the $x$ direction, and periodic boundary condition in the $y$ direction. The unit for $k_y$ is $\pi/a$, with $a$ the lattice spacing. The energy spectra  shown in (a, b) correspond to the Dirac type. The spectra shown in (c, d) correspond to the topological insulator type, where the in-gap edge states respecting the time-reversal symmetry are revealed. In this plot, we choose $t_\parallel =0.35 U_\parallel$, $t_\perp = 0.08 U_\parallel$, $U_\perp = U_\parallel$,  and the adjacent potential difference $\delta =0 $ and $=0.2 U_\parallel$ for (a,b) and (c,d), respectively.
}
\label{fig_topexcitations}
\end{figure}

We consider a two-dimensional optical lattice in presence of a chemical potential difference for adjacent sites, which induces the system into two non-equivalent sublattices, denoted as $A$ or $B$. We focus on topological Bogoliubov excitations of bosonic superfuids. Within the Bogoliubov approximation, the corresponding Hamiltonian can be rewritten as
\begin{eqnarray}
\hspace{-5mm}\boldsymbol{\mathcal{H}} \equiv
        \hspace{0mm} \left(\begin{array}{cc} \hspace{-2cm} H_0(k)+U_d(k) &   \hspace{-1.25cm} U_o(k) \\
                                             \hspace{0.cm}  U_o^\ast(-k)   \hspace{0.5cm} H^\ast_0(k)+U_d^\ast(-k)  \end{array}\hspace{0.25cm} \right) \hspace{-0.cm},	\nonumber
\end{eqnarray}
where $k$ denotes the momentum, $H_0(k)$ denotes the single-particle Hamiltonian in momentum space. We assume the ground state for the system is given by $\Phi_k= (\phi_{x\uparrow}, \phi_{x\downarrow}, \phi_{y\uparrow}, \phi_{y\downarrow})_{A,B}$ with particle number being $(N_{x\uparrow},N_{x\downarrow}, N_{y\uparrow}, N_{y\downarrow})_{A,B}$. We then drive the interaction terms in momentum space, and without loss of generality, we only write down the interaction part for sublattice $A$ (Note here that sublattice $A$ and $B$ are coupled). Here, we define the $ U_d({\bf k})$ and $ U_o({\bf k})$:
\begin{equation}
  U_d({k}) =
  \begin{pmatrix}
    M_{11} & M_{12} & M_{13} & M_{14} \\
    M_{21} & M_{22} & M_{23} & M_{24} \\
    M_{31} & M_{32} & M_{33} & M_{34} \\
    M_{41} & M_{42} & M_{43} & M_{44}
  \end{pmatrix}
, {\,\rm and \,\,}
  U_o({k}) =
  \begin{pmatrix}
    N_{11} & N_{12} & N_{13} & N_{14} \\
    N_{21} & N_{22} & N_{23} & N_{24} \\
    N_{31} & N_{32} & N_{33} & N_{34} \\
    N_{41} & N_{42} & N_{43} & N_{44}
  \end{pmatrix},
\end{equation}
where each element of the matrixs takes the following form:
\begin{subequations}
\begin{equation}
  \begin{split}
  M_{11}=& \frac{U_{\uparrow\downarrow} N_{x \downarrow} }{2N_u} (\phi^*_{x \downarrow} \phi_{x \downarrow} - \phi^*_{x \uparrow} \phi_{x \uparrow}\phi^*_{x \downarrow} \phi_{x \downarrow} - \Phi_x ) - \frac{U_{\uparrow\uparrow} N_{x \uparrow} }{2N_u}  (\phi^*_{x \uparrow} \phi_{x \uparrow}\phi^*_{x \uparrow} \phi_{x \uparrow} + \Phi_x) \\
  & + \frac{U_{\uparrow\uparrow}N_{x\uparrow}}{N_u}\phi^*_{x \uparrow} \phi_{x \uparrow}  - \frac{V_{\uparrow\downarrow} N_{y \downarrow}}{2 N_u} (\phi^*_{x \uparrow} \phi_{x \uparrow}\phi^*_{y \downarrow} \phi_{y \downarrow} + \Phi^\prime_{xy}) \\
  & -\frac{V_{\uparrow\uparrow} N_{y \uparrow}}{2N_u}  [\frac{1}{2} (\phi^*_{x \uparrow} \phi^*_{x \uparrow}\phi_{y \uparrow} \phi_{y \uparrow} + \phi^*_{y \uparrow} \phi^*_{y \uparrow}\phi_{x \uparrow} \phi_{x \uparrow}) + \Phi_{xy}]  + \frac{V_{\uparrow\downarrow}}{2 N_u}  N_{y \downarrow} \phi^*_{y \downarrow} \phi_{y \downarrow} \\
  & + \frac{V_{\uparrow\uparrow}}{ N_u}  N_{y \uparrow} \phi^*_{y \uparrow} \phi_{y \uparrow} - \frac{V_{\uparrow\uparrow}}{ N_u} N_{y \uparrow} (\phi^*_{x \uparrow} \phi_{x \uparrow}\phi^*_{y \uparrow} \phi_{y \uparrow}+\Phi^\prime_{xy}),
  \end{split}
\end{equation}
\begin{equation}
  M_{12}=  {\frac{U_{\uparrow\downarrow} \sqrt{ N_{x \uparrow} N_{x \downarrow}}  }{2N_u} \phi^*_{x \downarrow} \phi_{x \uparrow} } + {\frac{V_{\uparrow\downarrow}}{2 N_u}  \sqrt{N_{y \uparrow}N_{y \downarrow}} \phi^*_{y \downarrow} \phi_{y \uparrow} },
\end{equation}
\begin{equation}
\begin{split}
  M_{13} =& \frac{V_{\uparrow\uparrow} \sqrt{N_{x\uparrow}N_{y\uparrow}}}{N_u} \phi^*_{x \uparrow} \phi_{y \uparrow} + {\frac{V_{\uparrow\downarrow} \sqrt{N_{x\downarrow}N_{y\downarrow}}}{2N_u} \phi^*_{x \downarrow} \phi_{y \downarrow}} + {\frac{V_{\uparrow\uparrow} \sqrt{N_{x \uparrow} N_{y \uparrow}} }{2 N_u} \phi^*_{y \uparrow} \phi_{x \uparrow} } \\
  &+{\frac{V_{\uparrow\uparrow} \sqrt{N_{x \uparrow} N_{y \uparrow} }}{2 N_u} \phi^*_{y \uparrow} \phi_{x \uparrow} + \frac{V_{\uparrow\downarrow} \sqrt{N_{x \downarrow} N_{y \downarrow} }}{2 N_u} \phi^*_{y \downarrow} \phi_{x \downarrow}},
\end{split}
\end{equation}
\begin{equation}
  M_{14} = {\frac{V_{\uparrow\downarrow} \sqrt{N_{x\uparrow}N_{y\downarrow}}}{2N_u} \phi^*_{x \uparrow} \phi_{y \downarrow} } +  {\frac{V_{\uparrow\downarrow} \sqrt{N_{x \uparrow} N_{y \downarrow}} }{2 N_u} \phi^*_{y \downarrow} \phi_{x \uparrow}  },
\end{equation}
\begin{equation}
  M_{21} = {\frac{U_{\uparrow\downarrow} \sqrt{ N_{x \uparrow} N_{x \downarrow}}  }{2N_u} \phi^*_{x \uparrow} \phi_{x \downarrow}} + {\frac{V_{\uparrow\downarrow}}{2 N_u}  \sqrt{N_{y \uparrow}N_{y \downarrow}} \phi^*_{y \uparrow} \phi_{y \downarrow} },
\end{equation}
\begin{equation}
\begin{split}
  M_{22} =& \frac{U_{\uparrow\downarrow} N_{x \uparrow} }{2N_u} (\phi^*_{x \uparrow} \phi_{x \uparrow} - \phi^*_{x \uparrow} \phi_{x \uparrow}\phi^*_{x \downarrow} \phi_{x \downarrow}-\Phi_x) - \frac{U_{\downarrow\downarrow} N_{x \downarrow} }{2N_u} ( \phi^*_{x \downarrow} \phi_{x \downarrow}\phi^*_{x \downarrow} \phi_{x \downarrow}+\Phi_x) \\
  & + \frac{U_{\downarrow\downarrow}N_{x\downarrow}}{N_u}\phi^*_{x \downarrow} \phi_{x \downarrow}- \frac{V_{\uparrow\downarrow} N_{y \uparrow} }{2 N_u}(\phi^*_{x \downarrow} \phi_{x \downarrow}\phi^*_{y \uparrow} \phi_{y \uparrow} + \Phi^\prime_{xy})  \\
  & -\frac{V_{\downarrow\downarrow}N_{y \downarrow}}{2N_u}  [\frac{1}{2} (\phi^*_{x \downarrow} \phi^*_{x \downarrow}\phi_{y \downarrow} \phi_{y \downarrow} + \phi^*_{y \downarrow} \phi^*_{y \downarrow}\phi_{x \downarrow} \phi_{x \downarrow}) + \Phi_{xy}] + \frac{V_{\uparrow\downarrow}N_{y \uparrow}}{2 N_u}   \phi^*_{y \uparrow} \phi_{y \uparrow}\\
  & +  \frac{V_{\downarrow\downarrow}}{ N_u}  N_{y \downarrow} \phi^*_{y \downarrow} \phi_{y \downarrow} - \frac{V_{\downarrow\downarrow} N_{y \downarrow}}{ N_u} (\phi^*_{x \downarrow} \phi_{x \downarrow}\phi^*_{y \downarrow} \phi_{y \downarrow} + \Phi^\prime_{xy}),
\end{split}
\end{equation}
\begin{equation}
  M_{23} = {\frac{V_{\uparrow\downarrow} \sqrt{ N_{x\downarrow}N_{y\uparrow}}}{2N_u} \phi^*_{x \downarrow} \phi_{y \uparrow} } + {\frac{V_{\uparrow\downarrow} \sqrt{N_{x \downarrow} N_{y \uparrow}} }{2 N_u} \phi^*_{y \uparrow} \phi_{x \downarrow}},
\end{equation}
\begin{equation}
\begin{split}
  M_{24} =&  \frac{V_{\downarrow\downarrow} \sqrt{N_{x\downarrow}N_{y\downarrow}}}{N_u} \phi^*_{x \downarrow} \phi_{y \downarrow} + {\frac{V_{\uparrow\downarrow} \sqrt{N_{x\uparrow}N_{y\uparrow}}}{2 N_u} \phi^*_{x \uparrow} \phi_{y \uparrow}} + {\frac{V_{\downarrow\downarrow} \sqrt{N_{x \downarrow} N_{y \downarrow}} }{2 N_u} \phi^*_{y \downarrow} \phi_{x \downarrow}   }\\
  & + {\frac{V_{\uparrow\downarrow} \sqrt{N_{x \uparrow} N_{y \uparrow} }}{2 N_u} \phi^*_{y \uparrow} \phi_{x \uparrow} + \frac{V_{\downarrow\downarrow} \sqrt{N_{x \downarrow} N_{y \downarrow} }}{2 N_u} \phi^*_{y \downarrow} \phi_{x \downarrow}},
\end{split}
\end{equation}
\begin{equation}
\begin{split}
  M_{31} = &\frac{V_{\uparrow\uparrow} \sqrt{N_{x\uparrow}N_{y\uparrow}}}{N_u} \phi^*_{y \uparrow} \phi_{x \uparrow} + {\frac{V_{\uparrow\downarrow} \sqrt{N_{x\downarrow}N_{y\downarrow}}}{2N_u} \phi^*_{y \downarrow} \phi_{x \downarrow}} + {\frac{V_{\uparrow\uparrow} \sqrt{N_{x \uparrow} N_{y \uparrow}} }{2 N_u} \phi^*_{x \uparrow} \phi_{y \uparrow} } \\
  &+ {\frac{V_{\uparrow\uparrow} \sqrt{N_{x \uparrow} N_{y \uparrow} }}{2 N_u} \phi^*_{x \uparrow} \phi_{y \uparrow} + \frac{V_{\uparrow\downarrow} \sqrt{N_{x \downarrow} N_{y \downarrow} }}{2 N_u} \phi^*_{x \downarrow} \phi_{y \downarrow}},
\end{split}
\end{equation}
\begin{equation}
  M_{32} = {\frac{V_{\uparrow\downarrow} \sqrt{N_{x\uparrow}N_{y\downarrow}}}{2N_u} \phi^*_{y \uparrow} \phi_{x \downarrow} } + {\frac{V_{\uparrow\downarrow} \sqrt{N_{x \downarrow} N_{y \uparrow}} }{2 N_u} \phi^*_{x \downarrow} \phi_{y \uparrow}},
\end{equation}
\begin{equation}
\begin{split}
  M_{33} =& {\frac{U_{\uparrow\downarrow} N_{y \downarrow} }{2N_u} (\phi^*_{y \downarrow} \phi_{y \downarrow} - \phi^*_{y \uparrow} \phi_{y \uparrow}\phi^*_{y \downarrow} \phi_{y \downarrow}-\Phi_y) -  \frac{U_{\uparrow\uparrow} N_{y \uparrow} }{2N_u}  (\phi^*_{y \uparrow} \phi_{y \uparrow}\phi^*_{y \uparrow} \phi_{y \uparrow}} + \Phi_y)\\
  & +{\frac{U_{\uparrow\uparrow}N_{y\uparrow}}{N_u}\phi^*_{y \uparrow} \phi_{y \uparrow}} { - \frac{V_{\uparrow\downarrow} N_{x \downarrow}}{2 N_u}  (\phi^*_{x \downarrow} \phi_{x \downarrow}\phi^*_{y \uparrow} \phi_{y \uparrow}} +\Phi^\prime_{xy})\\
  & {-\frac{V_{\uparrow\uparrow} N_{x \uparrow}}{2N_u}   [\frac{1}{2} (\phi^*_{x \uparrow} \phi^*_{x \uparrow}\phi_{y \uparrow} \phi_{y \uparrow} + \phi^*_{y \uparrow} \phi^*_{y \uparrow}\phi_{x \uparrow} \phi_{x \uparrow}) }+ \Phi_{xy}] + { \frac{V_{\uparrow\downarrow}}{2 N_u}  N_{x \downarrow} \phi^*_{x \downarrow} \phi_{x \downarrow} } \\
  & +{\frac{V_{\uparrow\uparrow}}{ N_u}  N_{x \uparrow} \phi^*_{x \uparrow} \phi_{x \uparrow} }{- \frac{V_{\uparrow\uparrow}}{ N_u} N_{x \uparrow} (\phi^*_{x \uparrow} \phi_{x \uparrow}\phi^*_{y \uparrow} \phi_{y \uparrow}  } + \Phi^\prime_{xy}),
\end{split}
\end{equation}
\begin{equation}
  M_{34} = {\frac{U_{\uparrow\downarrow} \sqrt{ N_{y \uparrow} N_{y \downarrow}}  }{2N_u} \phi^*_{y \downarrow} \phi_{y \uparrow}} +  {\frac{V_{\uparrow\downarrow}}{2 N_u}  \sqrt{N_{x \uparrow}N_{x \downarrow}} \phi^*_{x \downarrow} \phi_{x \uparrow} }
\end{equation}
\begin{equation}
  M_{41} = {\frac{V_{\uparrow\downarrow} \sqrt{ N_{x\downarrow}N_{y\uparrow}}}{2N_u} \phi^*_{y \downarrow} \phi_{x \uparrow} } + {\frac{V_{\uparrow\downarrow} \sqrt{N_{x \uparrow} N_{y \downarrow}} }{2 N_u} \phi^*_{x \uparrow} \phi_{y \downarrow}  },
\end{equation}
\begin{equation}
\begin{split}
  M_{42} =& \frac{V_{\downarrow\downarrow} \sqrt{N_{x\downarrow}N_{y\downarrow}}}{N_u} \phi^*_{y \downarrow} \phi_{x \downarrow} + {\frac{V_{\uparrow\downarrow} \sqrt{N_{x\uparrow}N_{y\uparrow}}}{2 N_u} \phi^*_{y \uparrow} \phi_{x \uparrow}} + {\frac{V_{\downarrow\downarrow} \sqrt{N_{x \downarrow} N_{y \downarrow}} }{2 N_u} \phi^*_{x \downarrow} \phi_{y \downarrow}   } \\
  & + {\frac{V_{\uparrow\downarrow} \sqrt{N_{x \uparrow} N_{y \uparrow} }}{2 N_u} \phi^*_{x \uparrow} \phi_{y \uparrow} + \frac{V_{\downarrow\downarrow} \sqrt{N_{x \downarrow} N_{y \downarrow} }}{2 N_u} \phi^*_{x \downarrow} \phi_{y \downarrow}},
\end{split}
\end{equation}
\begin{equation}
  M_{43} = {\frac{U_{\uparrow\downarrow} \sqrt{ N_{y \uparrow} N_{y \downarrow}}  }{2N_u} \phi^*_{y \uparrow} \phi_{y \downarrow}} + {\frac{V_{\uparrow\downarrow}}{2 N_u}  \sqrt{N_{x \uparrow}N_{x \downarrow}} \phi^*_{x \uparrow} \phi_{x \downarrow} },
\end{equation}
\begin{equation}
\begin{split}
  M_{44} = &{\frac{U_{\uparrow\downarrow} N_{y \uparrow} }{2N_u} (\phi^*_{y \uparrow} \phi_{y \uparrow} - \phi^*_{y \uparrow} \phi_{y \uparrow}\phi^*_{y \downarrow} \phi_{y \downarrow} -\Phi_y) -  \frac{U_{\downarrow\downarrow} N_{y \downarrow} }{2N_u} (\phi^*_{y \downarrow} \phi_{y \downarrow}\phi^*_{y \downarrow} \phi_{y \downarrow} } +\Phi_{y}) \\
  & + {\frac{U_{\downarrow\downarrow}N_{y\downarrow}}{N_u}\phi^*_{y \downarrow} \phi_{y \downarrow}} - \frac{V_{\uparrow\downarrow} N_{x \uparrow} }{2 N_u} (\phi^*_{y \downarrow} \phi_{y \downarrow}\phi^*_{x \uparrow} \phi_{x \uparrow} + \Phi^\prime_{xy}) \\
  &{-\frac{V_{\downarrow\downarrow} N_{x \downarrow}}{2N_u}  [\frac{1}{2} (\phi^*_{x \downarrow} \phi^*_{x \downarrow}\phi_{y \downarrow} \phi_{y \downarrow} + \phi^*_{y \downarrow} \phi^*_{y \downarrow}\phi_{x \downarrow} \phi_{x \downarrow}) } +\Phi_{xy}] + {\frac{V_{\uparrow\downarrow}}{2 N_u}  N_{x \uparrow} \phi^*_{x \uparrow} \phi_{x \uparrow}  } \\
  & + { \frac{V_{\downarrow\downarrow}}{ N_u}  N_{x \downarrow} \phi^*_{x \downarrow} \phi_{x \downarrow} }{- \frac{V_{\downarrow\downarrow}}{ N_u} N_{x \downarrow} (\phi^*_{x \downarrow} \phi_{x \downarrow}\phi^*_{y \downarrow} \phi_{y \downarrow}  } +\Phi^\prime_{xy}),
\end{split}
\end{equation}
\end{subequations}
\begin{subequations}
\begin{equation}
  N_{11}= \frac{U_{\uparrow\uparrow}  N_{x \uparrow} }  {2N_u} \phi_{x \uparrow} \phi_{x \uparrow} + \frac{V_{\uparrow\uparrow} N_{y\uparrow}}{2N_u} \phi_{y \uparrow} \phi_{y \uparrow},
\end{equation}
\begin{equation}
  N_{12}= \frac{U_{\uparrow\downarrow} \sqrt{ N_{x \uparrow} N_{x \downarrow}}  }{2N_u} \phi_{x \uparrow} \phi_{x \downarrow} + \frac{V_{\uparrow\downarrow} \sqrt{N_{y\uparrow}N_{y\downarrow}}}{2N_u} \phi_{y \uparrow} \phi_{y \downarrow},
\end{equation}
\begin{equation}
  N_{13} = \frac{V_{\uparrow\uparrow} \sqrt{N_{x \uparrow} N_{y \uparrow}} }{2 N_u} \phi_{x \uparrow} \phi_{y \uparrow} + \frac{V_{\uparrow\uparrow} \sqrt{N_{x \uparrow} N_{y \uparrow} }}{2 N_u} \phi_{x \uparrow} \phi_{y \uparrow},
\end{equation}
\begin{equation}
  N_{14} = \frac{V_{\uparrow\downarrow} \sqrt{N_{x \uparrow} N_{y \downarrow}} }{2 N_u} \phi_{x \uparrow} \phi_{y \downarrow} + \frac{V_{\uparrow\downarrow} \sqrt{N_{x \downarrow} N_{y \uparrow} }}{2 N_u} \phi_{x \downarrow} \phi_{y \uparrow},
\end{equation}
\begin{equation}
  N_{21} = \frac{U_{\uparrow\downarrow} \sqrt{ N_{x \uparrow} N_{x \downarrow}}  }{2N_u} \phi_{x \uparrow} \phi_{x \downarrow} + \frac{V_{\uparrow\downarrow} \sqrt{N_{y\uparrow}N_{y\downarrow}}}{2N_u} \phi_{y \uparrow} \phi_{y \downarrow},
\end{equation}
\begin{equation}
  N_{22} = \frac{U_{\downarrow\downarrow}  N_{x \downarrow} }  {2N_u} \phi_{x \downarrow} \phi_{x \downarrow} + \frac{V_{\downarrow\downarrow} N_{y\downarrow}}{2N_u} \phi_{y \downarrow} \phi_{y \downarrow},
\end{equation}
\begin{equation}
  N_{23} = \frac{V_{\uparrow\downarrow} \sqrt{N_{x \downarrow} N_{y \uparrow}} }{2 N_u} \phi_{x \downarrow} \phi_{y \uparrow} + \frac{V_{\uparrow\downarrow} \sqrt{N_{x \uparrow} N_{y \downarrow} }}{2 N_u} \phi_{x \uparrow} \phi_{y \downarrow},
\end{equation}
\begin{equation}
  N_{24} = \frac{V_{\downarrow\downarrow} \sqrt{N_{x \downarrow} N_{y \downarrow}} }{2 N_u} \phi_{x \downarrow} \phi_{y \downarrow} + \frac{V_{\downarrow\downarrow} \sqrt{N_{x \downarrow} N_{y \downarrow} }}{2 N_u} \phi_{x \downarrow} \phi_{y \downarrow},
\end{equation}
\begin{equation}
  N_{31} = \frac{V_{\uparrow\uparrow} \sqrt{N_{x \uparrow} N_{y \uparrow}} }{2 N_u} \phi_{x \uparrow} \phi_{y \uparrow} + \frac{V_{\uparrow\uparrow} \sqrt{N_{x \uparrow} N_{y \uparrow} }}{2 N_u} \phi_{x \uparrow} \phi_{y \uparrow},
\end{equation}
\begin{equation}
  N_{32} = \frac{V_{\uparrow\downarrow} \sqrt{N_{x \downarrow} N_{y \uparrow}} }{2 N_u} \phi_{x \downarrow} \phi_{y \uparrow} + \frac{V_{\uparrow\downarrow} \sqrt{N_{x \uparrow} N_{y \downarrow} }}{2 N_u} \phi_{x \uparrow} \phi_{y \downarrow},
\end{equation}
\begin{equation}
  N_{33} = \frac{U_{\uparrow\uparrow}  N_{y \uparrow} }  {2N_u} \phi_{y \uparrow} \phi_{y \uparrow} + \frac{V_{\uparrow\uparrow} N_{x\uparrow}}{2N_u} \phi_{x \uparrow} \phi_{x \uparrow},
\end{equation}
\begin{equation}
  N_{34} = \frac{U_{\uparrow\downarrow} \sqrt{ N_{y \uparrow} N_{y \downarrow}}  }{2N_u} \phi_{y \uparrow} \phi_{y \downarrow} + \frac{V_{\uparrow\downarrow} \sqrt{N_{x\uparrow}N_{x\downarrow}}}{2N_u} \phi_{x \uparrow} \phi_{x \downarrow},
\end{equation}
\begin{equation}
  N_{41} = \frac{V_{\uparrow\downarrow} \sqrt{N_{x \uparrow} N_{y \downarrow}} }{2 N_u} \phi_{x \uparrow} \phi_{y \downarrow} + \frac{V_{\uparrow\downarrow} \sqrt{N_{x \downarrow} N_{y \uparrow} }}{2 N_u} \phi_{x \downarrow} \phi_{y \uparrow},
\end{equation}
\begin{equation}
  N_{42} = \frac{V_{\downarrow\downarrow} \sqrt{N_{x \downarrow} N_{y \downarrow}} }{2 N_u} \phi_{x \downarrow} \phi_{y \downarrow} + \frac{V_{\downarrow\downarrow} \sqrt{N_{x \downarrow} N_{y \downarrow} }}{2 N_u} \phi_{x \downarrow} \phi_{y \downarrow},
\end{equation}
\begin{equation}
  N_{43} = \frac{U_{\uparrow\downarrow} \sqrt{ N_{y \uparrow} N_{y \downarrow}}  }{2N_u} \phi_{y \uparrow} \phi_{y \downarrow} + \frac{V_{\uparrow\downarrow} \sqrt{N_{x\uparrow}N_{x\downarrow}}}{2N_u} \phi_{x \uparrow} \phi_{x \downarrow},
\end{equation}
\begin{equation}
  N_{44} = \frac{U_{\downarrow\downarrow}  N_{y \downarrow} }  {2N_u} \phi_{y \downarrow} \phi_{y \downarrow} + \frac{V_{\downarrow\downarrow} N_{x\downarrow}}{2N_u} \phi_{x \downarrow} \phi_{x \downarrow},
\end{equation}
\end{subequations}
where, $N_u$ is the number of total unit cells, $U_{\uparrow\uparrow}$, $U_{\downarrow\downarrow}$, $V_{\uparrow\uparrow}$ and $V_{\downarrow\downarrow}$ denote the intraspecies interactions ($U_{\uparrow\uparrow}=U_{\downarrow\downarrow}\equiv U_\parallel$, $V_{\uparrow\uparrow}=V_{\downarrow\downarrow}\equiv V_\parallel$, see the main text for the definition), and $U_{\uparrow\downarrow}$ and $V_{\uparrow\downarrow}$ denote the interspecies interactions ($U_{\uparrow\downarrow}\equiv U_\perp$ and $V_{\uparrow\downarrow}\equiv V_\perp$, see the main text for the definition). Here, we have defined $\Psi_\nu =  \phi^*_{\nu \sigma} \phi^*_{\nu \sigma'} \phi_{\nu \sigma} \phi_{\nu \sigma'}$, $\Psi_{\mu\nu} = \frac{1}{2}\sum_{\mu\neq\nu} \phi^*_{\mu \sigma} \phi^*_{\mu \sigma} \phi_{\nu \sigma} \phi_{\nu \sigma}$, and $\Psi'_{\mu\nu}= \phi^*_{\mu \sigma}  \phi_{\mu \sigma} \phi^*_{\nu \sigma'} \phi_{\nu \sigma'}$ for the adjacent sites in sublattice $B$.

Based on Bogoliubov approximation, we study the excitation spectra of the bipartite square lattice. To simulate a real bosonic system loaded into p-bands, we takes hopping terms between next-nearest neighboring orbitals into account, in addition to the nearest ones. The ratio between the next-nearest neighboring tunnelings and nearest ones is set to be ten percent, based on the bandstructure calculation in Ref.~\onlinecite{2018_Zhifang_arXiv}. We remark that including the weak next-nearest tunnelings do not affect the robustness of the SAI order except causing minor modification of the order parameter strength and the phase boundary. We do observe a nontrival topological excitations with the development of edge states in between bulk bands induced by the chemical potential difference between adjacent sites with $\delta \neq 0$, as shown in the Fig.~\ref{fig_topexcitations}. In the absence of the chemical difference $\delta=0$, we instead find Dirac excitations.

\section{bosonic dynamical mean-field theory}
To investigate quantum phases of binary mixtures of spinor Bose gases loaded into an optical lattice, described by Eq.~(1), we establish a bosonic version of dynamical mean-field theory, and implement a parallel code to tackle the six-spin system with a huge Hilbert space. As in fermionic dynamical mean field theory, the main idea of the bosonic dynamical mean field theory (BDMFT) approach is to map the quantum lattice problem with many degrees of freedom onto a single site - ``impurity site" - coupled self-consistently to a noninteracting bath~\cite{Appen_georges96}. The dynamics at the impurity site can thus be thought of as the interaction (hybridization) of this site with the bath. Note here that this method is exact for infinite dimensions, and is a reasonable approximation for high but finite dimensions.
More specifically, we apply real-space bosonic dynamical mean-field theory (RBDMFT) which provides a non-perturbative description of the many-body system both in three and two spatial dimensions (considered here).
 RBDMFT operates on a finite realization of a lattice by the Dyson equation:
\begin{equation}
[G^{-1}]_{ij} = [G_0^{-1}]_{ij} - \Sigma_{ij}
\end{equation}
where $i$ and $j$ denote single sites of the system, and $G_0$ is the non-interacting Green's function. The core assumption of RBDMFT, as with standard BDMFT, is that the self-energy is local but site-dependent, $\Sigma_{ij} = \Sigma_i\delta_{ij}$, which is capable of including the inhomogeneity of a lattice system as well as strong correlations between the atoms. To quantitatively determine the phase boundary in our system, we take a finite system and enforce periodic boundary conditions. We investigate system sizes up to $32\times32$ sites on the square lattice, and indeed observe a nearly universal phase boundary regardless of the system size.

\subsection{BDMFT equations}
In deriving the effective action, we consider the limit of a high but finite dimensional optical lattice, and use the cavity method~\cite{Appen_georges96} to derive self-consistency equations within BDMFT. In the following, we use the notation $t_{ij} = t_\parallel \delta_{\mu \nu} -t_\perp (1-\delta_{\mu \nu} )$ between sites $i$ and $j$ to shorten Hamiltonian (see Eq.~(1)). We also assume that the Wannier function of the lowest energy band $\omega_{\sigma}({\bf x}-{\bf x}_i)$ is well localized in the $i$th lattice site, where $\sigma$ denotes the spin state $\uparrow$ ($\downarrow$) in $p_x$- or $p_y$- orbitals. Expanding a field operator by Wannier functions of the lowest energy band, $\hat{\Phi}_{\sigma} ({\bf x}) = \sum_i b_{i\sigma} \omega_{\sigma}({\bf x}-{\bf x}_i)$, and then the effective action of the impurity site up to subleading order in $1/z$ is then expressed in the standard way~\cite{Appen_georges96, Appen_Byczuk_2008}, which is described by:
\begin{eqnarray}\label{eff_action}
S^{(0)}_\text{imp} &=& -\int_0^\beta \hspace{-0.2cm} d \tau d\tau' \sum_{\sigma\sigma'} \Bigg( \hspace{-0.1cm} \begin{array}{c} b^{(0)*}_{\sigma} (\tau)\quad b^{(0)}_{\sigma} (\tau) \end{array}\hspace{-0.1cm} \Bigg)^{\hspace{0.1cm}} \boldsymbol{\mathcal{G}}^{(0)-1}_{\sigma\sigma'}(\tau-\tau') \Bigg(\begin{array}{c} \hspace{-0.1cm} b^{(0)}_{\sigma'} (\tau')\\ b^{(0)*}_{\sigma'} (\tau') \end{array} \hspace{-0.1cm}\Bigg) 	 \\
&+& \int_0^\beta d\tau \left\{ \frac{U_0}{6}   \left[ :2n^2(\tau) : - : L_z^2(\tau) :  + :{\bf S} ^2(\tau): \right]
+\frac{U_2}{6} \left[ :n^2(\tau): - : {LS_z} ^2(\tau): + (:3S_z ^2(\tau): - :{\bf S} ^2(\tau):)  \right] \right\} \nonumber
.\label{action}
\end{eqnarray}
Here, we have defined the Weiss Green's function (being a $8\times8$ matrix),
\begin{eqnarray}
&&\hspace{-5mm}\boldsymbol{\mathcal{G}}^{(0)-1}_{\sigma\sigma'}(\tau-\tau') \equiv - \\
&&\hspace{-5mm} \left(\begin{array}{cc} \hspace{-0.1cm}
				(\partial_{\tau'}-\mu_\sigma)\delta_{\sigma \sigma'}+
\hspace{-0.25cm} \sum \limits_{\langle 0i\rangle,\langle 0j\rangle} \hspace{-0.25cm}
t_{ij}^2G_{\sigma \sigma', ij}^1 (\tau, \tau')
			&   \hspace{-0.25cm} \sum \limits_{\langle
0i\rangle,\langle 0j\rangle} \hspace{-0.25cm} t_{ij}^2   G^2_{\sigma\sigma', ij}(\tau, \tau') \\
			  \hspace{-0.25cm} \sum \limits_{\langle
0i\rangle,\langle 0j\rangle} \hspace{-0.25cm} t_{ij}^2 {G^2_{\sigma\sigma', ij}}^*(\tau', \tau)
	 & (-\partial_{\tau'}-\mu_\sigma)\delta_{\sigma \sigma'}+   \hspace{-0.25cm}
\sum \limits_{\langle 0i\rangle,\langle 0j\rangle} \hspace{-0.25cm} t_{ij}^2  G_{\sigma \sigma', ij}^1
(\tau', \tau)
			 \hspace{-0.1cm} \end{array}\right) \hspace{-0.15cm},	\nonumber
\end{eqnarray}
and introduced
\begin{equation}
\phi^{}_{i,\sigma}(\tau) \equiv \langle b_{i, \sigma} (\tau)
\rangle_0
\end{equation}
as the superfluid order parameters, and
\begin{eqnarray}
\hspace{-0.5cm}G_{\sigma\sigma', ij}^1 (\tau, \tau')\hspace{-0.2cm}&\ \equiv\ &\hspace{-0.2cm}- \langle b_{i, \sigma} (\tau) b_{j, \sigma'}^* (\tau')
\rangle_0 + \phi_{i, \sigma'} (\tau) \phi_{j, \sigma}^* (\tau'), \\
\hspace{-0.5cm}G_{\sigma \sigma', ij}^2 (\tau, \tau')\hspace{-0.2cm}&\ \equiv\ &\hspace{-0.2cm}- \langle b_{i, \sigma} (\tau) b_{j, \sigma'} (\tau')
\rangle_0 + \phi_{i, \sigma'} (\tau) \phi_{j, \sigma} (\tau')
\end{eqnarray}
as the diagonal and off-diagonal parts of the connected Green's functions, respectively, where $\langle \ldots \rangle_0$ denotes the expectation value in the cavity system (without the impurity site). Note here that $\boldsymbol{\mathcal{G}}^{(0)-1}_{\sigma\sigma'}(\tau-\tau')$ is a $8\times8$ matrix with $\sigma$ ($\sigma^\prime$) running over all the possible values for the spin state $\uparrow$ ($\downarrow$) in $p_x$- and $p_y$- orbitals.

\subsection{Anderson impurity model}
The most difficult step in the procedure discussed above is to find a solver for the effective action. However, one cannot do this analytically. To obtain BDMFT equations, it is better to return back to the Hamiltonian representation. We find that the local Hamiltonian is given by a bosonic Anderson impurity model
\begin{eqnarray}
\hat{H}^{(0)}_A &=& - \sum_\sigma  \Bigg( t_\sigma \Big(\phi^{(0)*}_{\sigma} \hat{b}^{(0)}_{\sigma} + {\rm H.c.} \Big) + \frac{U_0}{6} \sum_{\bf r}   \left[ : 2n^2 : - : L_z^2 :  + :{\bf S} ^2: \right]
+ \frac{U_2}{6}  \sum_{\bf r}  \left[ :n^2: -: {LS_z} ^2: + (:3S_z ^2: - :3{\bf S} ^2:)  \right] \Bigg) \nonumber
\\
&+& + \sum_{l}  \epsilon_l \hat{a}^\dagger_l\hat{a}_l + \sum_{l,\sigma} \Big( V_{\sigma,l} \hat{a}_l\hat{b}^{\dagger(0)}_{\sigma} + W_{\sigma,l} \hat{a}_l\hat{b}^{(0)}_{\sigma} + {\rm H.c.} \Big),
\end{eqnarray}
where the chemical potential and interaction term are directly inherited from the Hubbard
Hamiltonian. The bath of condensed bosons is represented by the Gutzwiller term with
superfluid order parameters $\phi^{(0)}_{\sigma}$ for each component of the two species. The bath of normal bosons is
described by a finite number of orbitals with creation operators $\hat{a}^\dagger_l$ and energies
$\epsilon_l$, where these orbitals are coupled to the impurity via normal-hopping amplitudes
$V_{\sigma, l}$ and anomalous-hopping amplitudes $W_{\sigma, l}$. The anomalous hopping terms are
needed to generate the off-diagonal elements of the hybridization function.

We now turn to the solution of the impurity model. The Anderson Hamiltonian can straightforwardly be implemented in the Fock basis, and the corresponding solution can be achieved by exact diagonalization (ED) of fermionic DMFT~\cite{Appen_Caffarel_1994, Appen_georges96}. After diagonalization, the local Green's function, which includes all the information about the bath, can be obtained from the eigenstates and eigenenergies
in the Lehmann-representation
\begin{eqnarray}
G_{\rm imp,\sigma \sigma'}^1 (i \omega_n) &=& \frac{1}{Z} \sum_{mn} \langle m | \hat b_\sigma | n\rangle \langle n | \hat b_{\sigma'}^\dagger | m \rangle \frac{e^{- \beta E_n} - e^{-\beta E_m}}{E_n - E_m + i \hbar \omega_n} + \beta \phi_\sigma \phi^\ast_{\sigma'} \\
G_{\rm imp,\sigma \sigma'}^2 (i \omega_n) &=& \frac{1}{Z} \sum_{mn} \langle m | \hat b_\sigma | n\rangle \langle n | \hat b_{\sigma'} | m \rangle \frac{e^{- \beta E_n} - e^{-\beta E_m}}{E_n - E_m + i \hbar \omega_n} + \beta \phi_\sigma \phi_{\sigma'}.
\end{eqnarray}

Integrating out the orbitals leads to the same effective action as in Eq. (2) in the main text, if the following
identification is made
\begin{eqnarray}
\boldsymbol{\Delta}_{\sigma\sigma'} (i\omega_n)  & \equiv & t^2 {\sum_{\langle 0i\rangle ,\langle 0j \rangle }}^\prime\mathbf G^{(0)}_{\sigma\sigma',ij}(i\omega_n),
\end{eqnarray}
where $\sum^\prime$ means summation only over the nearest neighbors of the "impurity site", and we have defined the hybridization functions:
\begin{eqnarray}\label{hybridization}
\Delta_{\sigma\sigma'}^1(i \omega_n) & \equiv & -\sum_l\Big(\frac{V_{\sigma,l}V^\ast_{\sigma',l}}{\epsilon_l-i\omega_n} + \frac{W^\ast_{\sigma,l}W_{\sigma',l}}{\epsilon_l+i\omega_n}\Big), \nonumber \\
\Delta_{\sigma\sigma'}^2(i \omega_n) & \equiv &  -\sum_l\Big(\frac{V_{\sigma,l}W^\ast_{\sigma',l}}{\epsilon_l-i\omega_n} +
\frac{W^\ast_{\sigma,l}V_{\sigma',l}}{\epsilon_l+i\omega_n}\Big).
\label{hybridization}
\end{eqnarray}
Here, we make the approximation that the lattice self-energy $\Sigma_{\rm lat, \sigma\sigma'}$ coincides with the impurity self-energy $\Sigma_{\rm imp, \sigma\sigma'}$, which is obtained from the local Dyson equation
\begin{equation}
 G^{-1}_{ii} (i \omega_n) =  \left( \begin{array}{cc}
                          i \omega_n +\mu + \Delta_{\sigma\sigma'}^1 - \Sigma_{i,\sigma\sigma'}^{11}& \Delta_{\sigma\sigma'}^2-\Sigma_{i,\sigma\sigma'}^{12} \\
                         \Delta_{\sigma\sigma'}^{2\ast}- \Sigma_{i,\sigma\sigma'}^{21} \ & -i \omega_n + \mu +\Delta_{\sigma\sigma'}^{1\ast} - \Sigma_{i,\sigma\sigma'}^{22}
                         \end{array}\right).
 \label{eq:localdysonequation}
\end{equation}
The real-space Dyson equation takes the following form:
\begin{equation}
 G^{-1}_{ij, \mathrm{latt}} (i \omega_n) =  \!\left( \begin{array}{cc}
                          \!\!\left(i \omega_n \!+\!\mu\!  -\! \Sigma_{i,\sigma\sigma'}^{11}\right)\delta_{ij}\!+ \!t_{ij}& \!\!-\Sigma_{i,\sigma\sigma'}^{12} \delta_{ij}\\
                         \!\!- \Sigma_{i,\sigma\sigma'}^{21} \delta_{ij}\ & \!\!\left(\!-\!i \omega_n \!+\! \mu \!-\! \Sigma_{i,\sigma\sigma'}^{22}\right)\delta_{ij}\!+\!t_{ij}\!\!
                         \end{array}\right),
                         \label{eq:realspacedyson}
\end{equation}
Finally, we need a criterion to set values of parameters $\epsilon_l$, $V_l$ and $W_l$. In practice, the self-consistency loop is solved as follows: starting from an initial choice for the Anderson parameters and the superfluid order parameter, the Anderson Hamiltonian is constructed in the Fock basis and diagonalized to obtain the eigenstates and eigenenergies. The eigenstates and energies allow us to calculate the impurity Green's functions, and then obtain the lattice Green's functions via Eq.~(\ref{eq:realspacedyson}). Subsequently, new Anderson parameters are obtained, by fitting the Anderson hybridization functions from Eq.~(\ref{hybridization}) to new hybridization functions obtained from the lattice Dyson equation, which is done by a conjugate gradient method. With this new Anderson parameters, the procedure is iterated until convergence is reached.

\clearpage


\end{widetext}

\bibliography{references}

\end{document}